\documentclass[%
reprint,
superscriptaddress,
showpacs,preprintnumbers,
 amsmath,amssymb,
aps,
pre,
floatfix
]{revtex4-1}

\usepackage{float}
\usepackage{graphicx}
\usepackage{dcolumn}
\usepackage{bm}
\usepackage[multidot]{grffile}
\usepackage[colorlinks=true,allcolors=blue]{hyperref}
\usepackage[]{tikz}

\usepackage{layouts}
\usepackage{enumerate} 

\begin{document}


\title{Stabilisation of dynamics of oscillatory systems by non-autonomous perturbation}

\author{Maxime Lucas}
\email{m.lucas@lancaster.ac.uk}
\affiliation{Department of Physics, Lancaster University, Lancaster LA1 4YB, United Kingdom}
\affiliation{INFN and CSDC, Dipartimento di Fisica e Astronomia, Universit{\`a} di Firenze, 50019 Sesto Fiorentino, Firenze, Italy}

\author{Julian Newman}
\email{j.newman1@lancaster.ac.uk}
\affiliation{Department of Physics, Lancaster University, Lancaster LA1 4YB, United Kingdom}

\author{Aneta Stefanovska}
\email{aneta@lancaster.ac.uk}
\affiliation{Department of Physics, Lancaster University, Lancaster LA1 4YB, United Kingdom}

\date{\today}

\begin{abstract}
Synchronisation and stability under periodic oscillatory driving are well-understood, but little is known about the effects of aperiodic driving, despite its abundance in nature. Here, we consider oscillators subject to driving with slowly varying frequency, and investigate both short-term and long-term stability properties. For a phase oscillator, we find that, counter-intuitively, such variation is guaranteed to enlarge the Arnold tongue in parameter space. Using analytical and numerical methods that provide information on time-variable dynamical properties, we find that the growth of the Arnold tongue is specifically due to the growth of a region of intermittent synchronisation where trajectories alternate between short-term stability and short-term neutral stability, giving rise to stability on average. We also present examples of higher-dimensional nonlinear oscillators where a similar stabilisation phenomenon is numerically observed. Our findings help support the case that in general, deterministic non-autonomous perturbation is a very good candidate for stabilising complex dynamics.
\end{abstract}

\maketitle

\section{Introduction}

Complex oscillatory dynamics abounds in nature. Despite many real-life examples exhibiting stable oscillations with a time-varying frequency (\emph{e.g.}, \cite{Stefanovska2000, hramov2006,lancaster2016}), little is known theoretically about the properties of this type of behaviour. This kind of oscillation requires aperiodic external driving, making the system non-autonomous by nature \cite{kloeden2011}, such that most of the traditional analytical methods are unusable or insufficient. The case of periodic forcing with a constant frequency, which has been extensively investigated to date, is often too simplistic to account for reality.

Closely linked to the concept of stability is the concept of synchronisation. Synchronisation phenomena in the sciences, including phase synchronisation of complex oscillators, have drawn much attention over the last decades \cite{pikovsky2003}. Different mechanisms for achieving phase synchronisation, such as synchronisation by periodic forcing \cite{VanderPol1927}, by noise \cite{Stratonovich1967}, and by quasi-periodic forcing \cite{feudel2006}, as well as synchronisation of chaotic oscillators \cite{pecora1990, Rosenblum1996}, and networks of oscillators \cite{Kuramoto1975}, have been considered. At the same time that theoretical interest in synchronisation phenomena has been growing, real-life applications have been found to be prevalent in many diverse aspects of nature \cite{goldbeter1997,Glass2001,Dattani2017}, including circadian rhythms \cite{leloup1999, Hafner2012}, cardio-respiratory dynamics \cite{Stefanovska2000, Mcguinness2004, suprunenko2013}, metabolic oscillations \cite{lancaster2016}, the brain \cite{Tass:98,Varela2001}, and climate dynamics \cite{desaedeleer2013}. But once again, little is understood theoretically about synchronisation in the context of variable-frequency oscillations.

So then, there is a need for extensive and ongoing study of dynamical behaviour under deterministic oscillatory driving with a time-varying frequency, to fill the gap between the existing theory of deterministically driven systems where constant frequency is typically assumed and the statistical theory of systems driven by noise. In this manuscript, we present three major contributions to the field of interacting non-autonomous systems: Firstly, we present notions of stability, synchronisation and instantaneous frequency entrainment in the non-autonomous setting, and the relationships between these concepts; and we investigate these concepts for the simplest example of a phase oscillator subject to driving with slowly time-varying frequency. In so doing, we enable the notion of chronotaxicity \cite{suprunenko2013, suprunenko2014a, suprunenko2014b} to be broadened beyond its current description, and we compare the stability properties in this setting with the traditional settings of fixed-frequency driving on the one side and driving by stationary noise on the other. Secondly, we introduce an approach to analysing time-dependent dynamical stability from a time-series consisting of time-localised Lyapunov exponents (LE), that is, finite-time Lyapunov exponents (FTLE) taken over a time-window with a moving centre. By contrast, typically, dynamical stability is assessed only in terms of time-averaged stability, for example by the asymptotic LE \cite{pikovsky2016}. Thirdly, numerically and analytically, we show enlargement of the stability region in parameter space for the phase oscillator subject to driving with slowly varying frequency, and we show that this growth is specifically due to the growth of a subregion characterised by intermittent synchronisation where the time-localised dynamical stability is varying. While we show that slow modulation of the driving frequency guarantees enlargement of the stability region for one-dimensional phase oscillators, we also show numerically that the same phenomenon can readily occur in more general oscillatory systems. This mathematical phenomenon of stabilisation has two major practical implications: (i)~deterministically varying the frequency of external driving could be implemented as a means of inducing stability in complex systems, and (ii)~dynamical systems where stability is induced by deterministic frequency variation are an excellent candidate for modelling living systems, which are highly complex and yet usually operate stably within a time-varying environment.

The few existing pioneering studies of stability and synchronisation with time-varying frequency of oscillations have considered a simple case of linearly growing frequency \cite{garcia2008}, more general frequency that is slowly varied (as in our present paper) with particular application to both linearly growing frequency and low-pass-filtered noise \cite{jensen2002}, networks of coupled oscillators \cite{petkoski2012}, and a case of two interacting oscillators, each with the same form of time-varying frequency \cite{suprunenko2013, suprunenko2014b}. In the last-mentioned studies, the notion of chronotaxicity was introduced, to indicate a type of synchronisation specifically characteristic of non-autonomous interacting oscillators. The central idea in \cite{jensen2002} (also relevant to the analytical approach in the work presented here) is that under driving of sufficiently slowly varying frequency, the phenomenon of synchronisation by common driving can be investigated in terms of the presence of a stable equilibrium for the instantaneous vector field; the goal of \cite{jensen2002} is to identify what qualifies as slow variation. Time-series analysis methods have also been developed to resolve in time the dynamical characteristics of time-varying-frequency oscillators (\emph{e.g.}, wavelet-based spectrum, coherence and bispectrum \cite{clemson2014} as well as Bayesian inference of coupling functions \cite{stankovski2017}) rather than analyse them in a statistical sense (\emph{e.g.}\ calculating power-spectrum density) and thereby miss noteworthy time-dependent dynamical features. Normal forms of different types of non-autonomous bifurcations have  also been investigated \cite{Ashwin2011,Anagnostopoulou2012,Perryman2014}, and safety criteria were derived for aperiodically forced systems \cite{Bishnani2003}. 

However, this present work is the first investigation of non-autonomously driven oscillators showing growth of the stability region in parameter space, where this growth is due to time-variability without the need for statistical phenomena as in noisy models.

The paper is organised as follows. In Sec.~\ref{sec:one-dimensional}, we introduce a simple one-dimensional phase oscillator model. We then provide an explanation of notions of synchronisation and stability for non-autonomous systems, followed by a theoretical analysis of the one-dimensional model, showing the enlargement of the stability region, as well as the birth of an intermediate region of intermittent synchronisation. We illustrate these phenomena with numerical results for both long-time and short-time behaviour. In particular, in Sec.~\ref{sec:noise}, we discuss the relationship between the deterministic system considered here and the analogous case with noisy driving as considered in previous works. In Sec.~\ref{sec:higher-dimensional}, we illustrate the stabilisation phenomenon numerically in higher-dimensional systems, and argue that it is of general importance. Finally, in Sec.~\ref{sec:discussion} we discuss the results, and in Sec.~\ref{sec:summary} we provide a brief summary.


\section{One-dimensional case}
\label{sec:one-dimensional}

\subsection{Model}

The system we consider is a driven phase oscillator of the form
\begin{equation}
\dot{\theta} = \omega_0 + \gamma \sin (\theta - \theta_1 (t)),
\label{eq:system_polar_1}
\end{equation}
where the driving has strength $\gamma$, phase $\theta_1(t)$, and a time-varying frequency
\begin{equation}
\dot{\theta_1} = \omega_1 (1 + k f(\omega_m t)),
\label{eq:forcing_1_non-stationary}
\end{equation}
where $\omega_1$ is the non-modulated driving frequency, $f$ is a bounded function, and $\omega_m$ and $k$ are the modulation frequency and relative amplitude, respectively. The phase oscillator $\theta$ may represent, for example, the phase on the stable limit cycle of an oscillator satisfying
\begin{equation}
\left\{ \begin{array}{r c l}
\dot{r} & = & \epsilon(r_p-r)r \\
\dot{\theta} & = & \omega_0 + \gamma \sin (\theta - \theta_1 (t))
\end{array} \right.
\label{eq:full_osc}
\end{equation}
where $r_p$ is the amplitude of the limit cycle and $\epsilon$ is the restoring constant.

The unforced system \eqref{eq:system_polar_1} with $\gamma = 0$ is a typical autonomous phase oscillator \cite{strogatz1994}, and hence its phase is neutrally stable (zero LE).

In the forced system, \textit{i.e.}\ $\gamma \neq 0$, the traditional constant-frequency forcing case is recovered for $k=0$. In this case, depending on the parameters, the system lies in one of two regimes: either $1:1$~synchronisation (negative LE), or neutral stability (zero LE). The condition for synchronisation, $\gamma > | \Delta \omega |$, with the frequency mismatch $\Delta \omega = \omega_0 - \omega_1$, is derived analytically \cite{pikovsky2003} by requiring that the equation for the phase difference
\begin{equation}
\dot{\psi} = \Delta \omega + \gamma \sin \psi
\label{eq:auton_pd}
\end{equation}
has a stable fixed point. This condition for synchronisation corresponds to a so-called Arnold-tongue \cite{Arnold1983} in $(\gamma, \omega_0)$-parameter space. This Arnold tongue can be seen in Fig.~\ref{fig:diagram_lyap}(a) as the region appearing in shades of blue, corresponding to negative values of the numerically computed LE. Since \eqref{eq:auton_pd} is an autonomous differential equation, the numerical LE computed over a long time wil approximate well the asymptotic LE, except possibly when the parameters lie extremely close to the border of the Arnold tongue.

For $k \neq 0$, the equation for the phase difference is now the \textit{non-autonomous} equation
\begin{equation}
\dot{\psi} = \Delta \omega (t) + \gamma \sin \psi,
\label{eq:rotating_frame}
\end{equation}
with frequency mismatch $\Delta \omega (t) =\omega_0 - \omega_1 (1 + k f(\omega_m t))$. Throughout this paper, we assume that the modulation is much slower than the dynamics of the system, \textit{i.e.}\ $\omega_m$ is very small.

\subsection{Synchronisation in autonomous and non-autonomous systems}
\label{sec:nonaut}

Suppose an oscillatory system $\theta$ with no internal time-dependence is subject to driving from another oscillator $\theta_1$ with fixed frequency different from the natural frequency of $\theta$. We say that \emph{$\theta$ is synchronised to $\theta_1$} if over time, the trajectory of $\theta$ loses memory of its precise initial phase and instead follows a periodic behaviour whose period is a rational multiple $\frac{n}{m}$ of the period of $\theta_1$. In this case, we say that $\theta_1$ \emph{entrains} the frequency of $\theta$, and we describe the synchronisation as $n:m$~synchronisation. This implies in particular that the difference in unwrapped phase between an $n$-fold cycle of $\theta_1$ and an $m$-fold cycle of $\theta$ stays bounded over all time---a phenomenon referred to as \emph{phase-locking} between $\theta$ and $\theta_1$.

The particular phenomenon that $\theta(t)$ loses memory of its initial phase is called \emph{phase stability}. If $\theta$ is a phase oscillator (as in our model), then phase stability can be assessed in terms of the sign of Lyapunov exponent associated to the trajectory $\theta(t)$: a negative LE indicates stability. Stability inherently implies resilience against the effects of other possible perturbations not accounted for in the model. When the phase of a driven oscillator is stabilised by a fixed-frequency driving oscillator, typically this implies $n:m$~synchronisation for some integers $n$ and $m$; we emphasise that this statement is specific to fixed-frequency driving. For $n:m$~synchronisation where (in lowest terms) $n \geq 2$, it is not possible for $\theta(t)$ to lose \emph{all} memory of its initial phase: for any $1 \leq i \leq n-1$, delaying the initial phase by a suitable amount will delay the phase of the eventual periodic motion by $\frac{i}{n}$. However, in the case of $1:m$~synchronisation, it is possible for $\theta$ eventually to lose all memory of its initial phase; in this case, we say that the phase is \emph{globally stable}.

Synchronisation has also been investigated in the context of systems driven by noise, such as zero-mean Gaussian white noise or a pulse train with i.i.d.\ consecutive waiting times \cite{toral2001,pikovsky2003,pikovskii1984}. For an oscillator $\theta$ driven by such noise, one does not have a notion of $n:m$~synchronisation between this driven oscillator and the noise $\xi$ driving the oscillator. This is because, even if the noise is stationary noise, any one realisation of the noise does not have a deterministic periodic behaviour. As described in \cite[Section~15.2]{pikovsky2003}, instead of defining synchronisation in terms of ``phase locking'', one can think of synchronisation here as meaning that over time, $\theta$ loses memory of its precise initial state and instead follows \emph{some path} that is determined by the realisation of the noise---but since the noise itself has no deterministic regular behaviour, this phenomenon can only be physically manifested as \emph{synchronisation by common noise} between copies of $\theta$.

Synchronisation by common noise is a particular case of the phenomenon of synchronisation by common external driving (which may be noisy or deterministic): Suppose we have an array $\theta^1,\ldots,\theta^n$ of self-sustained oscillators whose internal dynamics are described by exactly the same system $\dot{\theta^i}=f(\theta^i)$, where $f$ does not depend on $i$; and no \emph{direct} coupling is introduced between these oscillators, but instead all these oscillators are \emph{simultaneously subject} to driving from the same external driver $p(t)$ (which could be noisy or deterministic). Thus the $n$ driven oscillators are now \emph{indirectly} coupled, and it may happen that as a result of this indirect coupling, over time the trajectories of $\theta^i$ lose memory of their initial states and instead follow the same path as each other. This may be viewed as a kind of perfectly instantaneous $1:1$~synchronisation between the driven oscillators.

The relationship between the above concepts is as follows. For a self-sustained phase oscillator $\theta$ subject to driving by an external driver $p(t)$, the following statements are equivalent:
\begin{enumerate}[(i)]
\item the trajectory of $\theta$ is globally stable;
\item the trajectory of $\theta$ eventually follows some path that is determined by $p(t)$ independently of the initial phase of $\theta$;
\item any array of identical copies of $\theta$ is synchronised when the indirect coupling of common driving by $p(t)$ is introduced;
\end{enumerate}
and in the case that $p(t)$ is a fixed-frequency deterministic oscillator, these typically imply that:
\begin{enumerate}[(iv)]
\item there is $1:m$~synchronisation between $p$ and $\theta$ for some $m$.
\end{enumerate}

The physical interpretation of the implication (ii)$\Rightarrow$(i) is that the driving $p(t)$ causes $\theta$ to become resilient in its course of following the path laid out by $p(t)$, although as we shall see, this resilience may only be intermittent. Such driving-induced resilience may play an important role in many real-world systems that exhibit remarkable stability in the face of continuous environmental perturbations.

In our model, if $k \neq 0$ then the driving is a deterministic oscillator $\theta_1$ with non-fixed frequency. Hence, it will be useful for us to discuss notions of synchronisation for oscillators subject to deterministic oscillatory driving with time-varying frequency. Such driving shares in common with fixed-frequency driving that it is deterministic and oscillatory, and it shares in common with noisy driving that it does not possess a phase which proceeds in cycles of a fixed period. Therefore, on the one hand, as with noisy driving, it is not clear that one can correctly define a notion of $n:m$~frequency entrainment, though the slightly weaker phenomenon of $n:m$ phase-locking can still occur; nonetheless, as in \cite{jensen2002}, one can still consider the question of whether identical copies of the driven oscillator are caused to synchronise by simultaneous driving from the driving oscillator.

Having stated that $n:m$~frequency entrainment is difficult to define in our setting, let us now highlight our slow variation assumption. Under this assumption, one can define a notion of \emph{instantaneous frequency entrainment}. In general, if a pair of phase oscillators $\theta,\theta_1$ is governed by a \emph{non-autonomous} differential equation
\begin{equation}
\left\{ \begin{array}{r c l}
\dot{\theta_1} & = & f_1(t,\theta_1) \\
\dot{\theta} & = & f_2(t,\theta_1,\theta)
\end{array} \right.
\label{eq:general_non}
\end{equation}
and it is assumed that $f_1(t,\cdot),f_2(t,\cdot,\cdot)$ vary slowly with $t$, then we can say that there is frequency entrainment at time $t$ if the solution of the associated autonomous differential equation
\begin{equation}
\left\{ \begin{array}{r c l}
\tfrac{d}{ds}\theta_1(s) & = & f_1(t,\theta_1(s)) \\
\tfrac{d}{ds}\theta(s) & = & f_2(t,\theta_1(s),\theta(s))
\end{array} \right.
\label{eq:assoc_auton}
\end{equation}
exhibits frequency entrainment. In the case of our model, at any time $t$, there is instantaneous $1:1$~frequency entrainment between $\theta$ and $\theta_1$ if and only if the differential equation $\frac{d}{ds}\psi(s) = \Delta \omega(t) + \gamma \sin \psi(s)$ has a stable fixed point [compare with Eq.~\eqref{eq:auton_pd}].

Just as negative Lyapunov exponents are connected with the presence of frequency entrainment for fixed-frequency driving, so likewise instantaneous frequency entrainment will typically be connected with negative \emph{finite-time} Lyapunov exponents defined over a suitable time-window. Finite-time Lyapunov exponents are a measure of stability over finite time-scales. We will use the term \emph{time-localised Lyapunov exponent} to refer to FTLE taken over a sliding time-window $[t,t+\tau]$ which slides along with time $t$. By contrast, we will use the term \emph{long-term Lyapunov exponent} to refer to a Lyapunov exponent taken over a long time-interval $[0,T]$; when clear from the context, we will sometimes drop the word ``long-term''. Technically, a long-term LE is still a finite-time Lyapunov exponent, but it plays a similar role to asymptotic LE for autonomous systems. Asymptotic LE need not exist for non-autonomous systems---indeed, non-autonomous systems need not even be well-defined over infinite-time. But moreover, even if they can be defined, asymptotic LE may not necessarily be physically relevant for the limited time-scales on which a system is considered in practice.

\subsection{Theoretical analysis}
\label{sec:theoretical}

\begin{figure}[t]
	\centering
	\includegraphics[width=\linewidth]{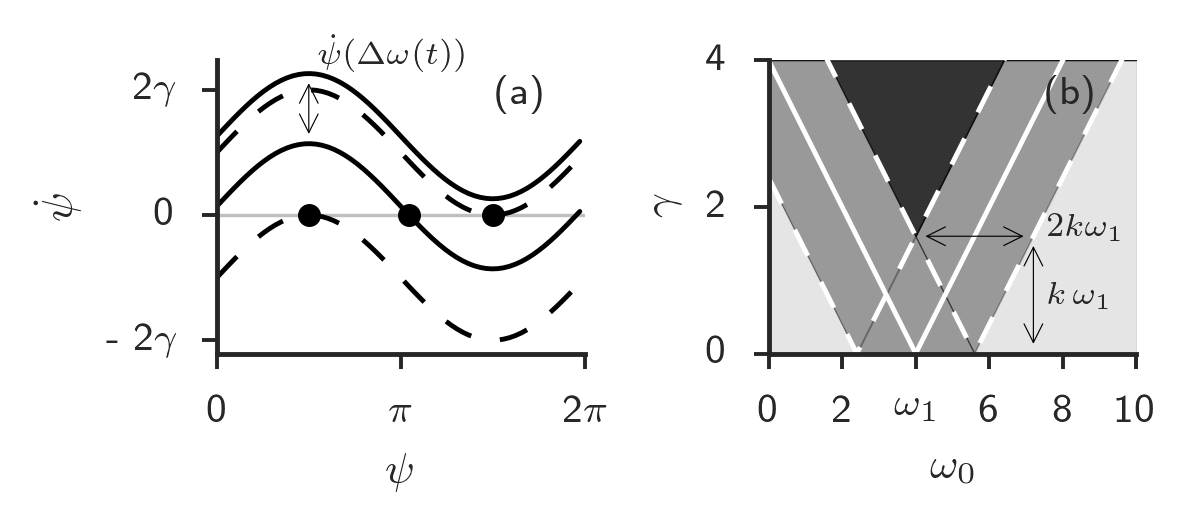}
	\caption{(a) Time-varying existence of the attracting point of Eq.~\eqref{eq:rotating_frame}. The $(\Delta \omega (t))$-dependent curve of $\dot{\psi}$ against $\psi$ moves up and down over time, as indicated by the two solid lines representing where the curve could be at two different instants in time. When this curve lies between the dashed lines, the system has an attracting point, and otherwise, not. (b)~Phase diagram showing three regimes. Light (region I in the text), medium (region III), and dark grey (region II) show where the system is never synchronising, intermittently synchronising, and always synchronising, respectively. Solid white curves show the border between synchronisation and non-synchronisation for if $f(\cdot)$ is set to $0$; and white dashed curves show the border between synchronisation and non-synchronisation for if $f(\cdot)$ is set to $\pm 1$. When $k$ is increased, regions~I and II decrease while region~III increases; hence in particular, the Arnold tongue consisting of the union of regions~II and III increases.}
	\label{fig:k-fattening}
\end{figure}

In contrast to the autonomous case, the existence of an attracting equilibrium point for the vector field on the right-hand side of Eq.~\eqref{eq:rotating_frame} (regarded as a function of $\psi$) can change with time $t$; as shown in Fig.~\ref{fig:k-fattening}(a), for a sinusoidal modulation $f(\cdot)=\sin(\cdot)$, if $k$ is large enough then within each modulation period the vector field undergoes two saddle-node bifurcations. Since we assume that the modulation is much slower than the dynamics of the system, the system adiabatically follows the moving attracting point $\psi_{\text{slow}}(t) = \pi - \arcsin[- \Delta \omega (t) / \gamma]$ for Eq.~\eqref{eq:rotating_frame}, when it exists. (A more technically precise description of how $\omega_m$ needs to compare with the values of other parameters in order to qualify as ``small'' for the purposes of this adiabatic approach can be found in \cite{jensen2002}.) On faster time-scales, one could view $\Delta \omega (t)$ as approximately constant and consider the stable point in the quasi-stationary limit.

Following the idea that solutions follow the moving attracting point $\psi_{\text{slow}}$ when it exists, we derive three regions, with qualitative features corresponding to the following conditions on Eq.~\eqref{eq:rotating_frame}: (I) no existence of the attracting point at any time $t$, (II) existence of the attracting point for all $t$, and (III) alternation over time between the existence and non-existence of the attracting point. If we assume that $f(\cdot)$ varies throughout the interval $[-1,1]$, then these conditions on the parameters are precisely
\begin{align}
\text{(I): } \quad & \gamma \leq |\omega_0 - \omega_1| - \omega_1 k, \label{eq:qualitative_regions1} \\
\text{(II): } \quad & \gamma \geq |\omega_0 - \omega_1| + \omega_1 k, \label{eq:qualitative_regions2} \\
\text{(III): } \quad & |\omega_0 - \omega_1| - \omega_1 k \leq \gamma \leq |\omega_0 - \omega_1| + \omega_1 k,
\label{eq:qualitative_regions3}
\end{align}
as illustrated in Fig.~\ref{fig:k-fattening}(b). In region~I, the slow variation assumption imlies that solutions behave similarly to the neutrally stable regime of the fixed-frequency-driving system; solutions of \eqref{eq:system_polar_1} or \eqref{eq:rotating_frame} will exhibit neutral stability, with a long-term Lyapunov exponent that is essentially zero. In region~II, the attracting point exists at all times, and attracts solutions starting from throughout the circle to itself; thus, the driven oscillator $\theta$ is globally stable, losing memory of its initial state and following the motion of $\theta_1(t)+\psi_{\text{slow}}(t)$. In particular, long-term LEs will be negative. There is instantaneous $1:1$~frequency entrainment at all times; moreover, the attracting point moves within a bounded arc of the circle, and thus we have $1:1$ phase-locking between $\theta$ and $\theta_1$.

In region~III, the attracting point exists at some times but not other times. We refer to the epochs during which the attracting point exists as stable epochs; the remain epochs are epochs of neutrally stable dynamics. During the stable epochs, solutions from throughout the circle are attracted to the attracting point. While following the attracting point, these solutions pick up a negative contribution to the Lyapunov exponent, due to the gradient of the instantaneous vector field being itself negative at the attracting point; and then during each of the epochs of neutral stability, the solutions receive zero net contribution to the Lyapunov exponent, meaning that overall, as in \cite{jensen2002}, long-term LE are negative and the solutions remain synchronised with each other over all time. There is instantaneous $1:1$~frequency entrainment during the stable epochs but no instantaneous frequency entrainment during the epochs of neutral stability; we will refer to this phenomenon as \emph{intermittent synchronisation}. Overall, we do not have phase-locking between $\theta$ and $\theta_1$. However, unlike in the case of fixed-frequency driving, synchrony of an array of identical copies of $\theta$ [represented as different simultaneous solutions of Eq.~\eqref{eq:system_polar_1}] is achieved and endures (even through the epochs of neutral stability) in the absence of a phase-locking mechanism. In other words, there does not need to be a phase-locking mechanism in place in order for the driving $\theta_1(t)$ to cause $\theta$ to lose all memory of its initial condition and follow a path determined by the evolution of $\theta_1(t)$.

Let us mention that there will be some very small subregions of region~III where in theory, if one waits long enough, $\theta$ will come close to the instantaneous repeller around the start of a stable epoch \cite{Guckenheimer2001} and thus receive a positive contribution to the LE, such that the reasoning here and in \cite{jensen2002} can eventually break down and the asymptotic LE (if $f(\cdot)$ is defined \emph{ad infinitum}) could even be zero. However, this theoretical phenomenon is unlikely to manifest in practice, due to the precision of fine-tuning of $f(\cdot)$ required for the phenomenon to occur, combined with the unphysical length of time that one is likely to have to wait for the phenomenon to take place. Indeed, no such phenomenon is ever observed in our numerics.

So then, in analogy to the case of fixed-frequency driving, we define the \emph{Arnold tongue} as being the union of region~II and region~III, that is, the total region where the long-term LE will be negative.

%
%

From Eq.~\eqref{eq:qualitative_regions3}, the role of the modulation amplitude $k$ here is clear: as $k$ increases from $0$, regions I and II decrease in size (although still extending infinitely), being symmetrically pushed back by the appearance and growth of region III, such that overall, the Arnold tongue is enlarged. In other words, increasing modulation amplitude induces stability.

\subsection{Numerical results}
\label{sec:numerics}

\begin{figure*}[hbt]
	\centering
	\includegraphics[width=0.98\linewidth]{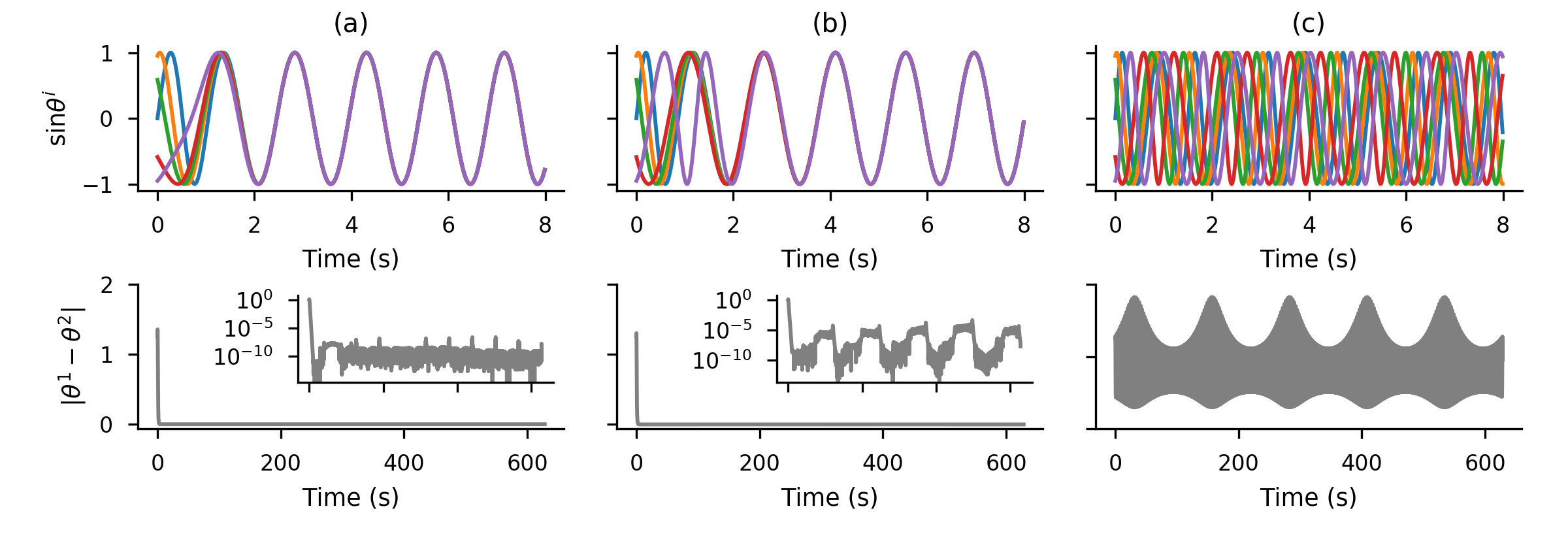}
	\caption{Trajectories of five solutions $\theta^1(t),\ldots,\theta^5(t)$ of Eq.~\eqref{eq:system_polar_1}, with initial conditions $\theta^i(0)=\frac{i-1}{5}.2\pi$, subject to the same driving $\theta_1(t)$. Here, $\gamma=2.5$ and $k=0.4$. In (a), $\omega_0=4$, and so the system is in region~II according to Eq.~\eqref{eq:qualitative_regions2}; in (b), $\omega_0=6$, and so the system is in region~III according to Eq.~\eqref{eq:qualitative_regions3}; in (c), $\omega_0=9$, and so the system is in region~I according to Eq.~\eqref{eq:qualitative_regions1}. In each case, the upper plot shows the first 8~seconds of the sine of the five trajectories, while the lower plot shows the distance between $\theta^1(t)$ and $\theta^2(t)$ over about the first 630~seconds (more precisely, 5~cycles of the frequency modulation); in (a) and (b), the inner graph shows the same information on a logarithmic scale. In (a) and (b), the system lies within the Arnold tongue as described in Sec.~\ref{sec:theoretical}, and the five trajectories are observed to synchronise and to remain in synchrony; by contrast, in (c), the system does not lie within the Arnold tongue, and no synchronisation is observed.}
	\label{fig:convergence}
\end{figure*}

\begin{figure}[h]
	\centering
	\includegraphics[width=\linewidth]{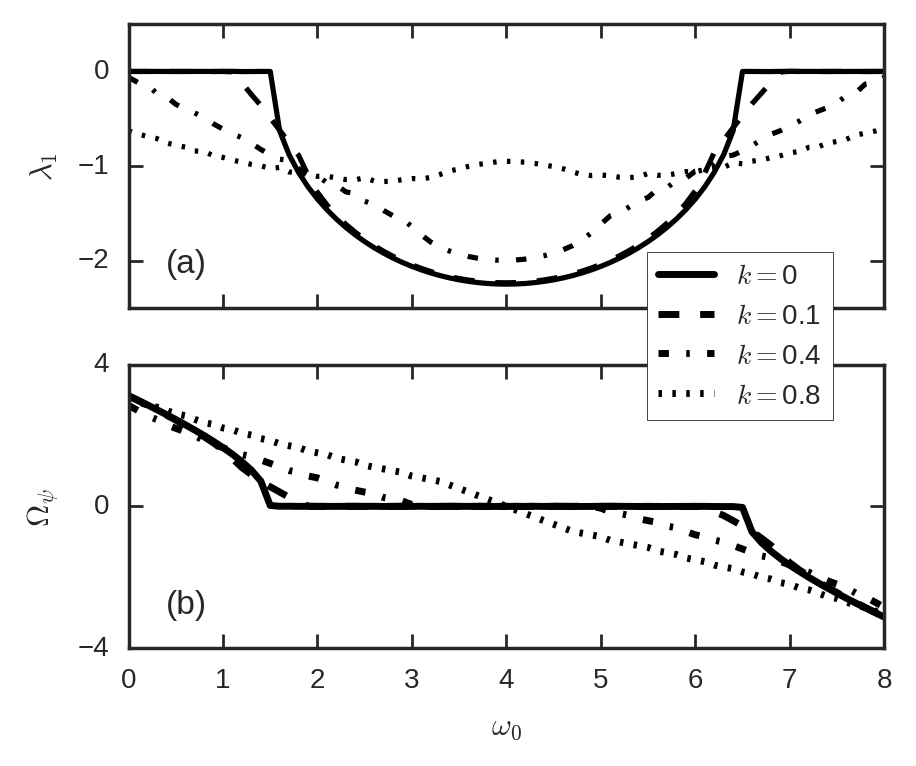}
	\caption{Numerically obtained time-averaged stability properties for a trajectory of Eq.~\eqref{eq:system_polar_1} starting at $\theta(0)=0$, computed over 10 cycles of the frequency modulation (about 1260~s). (a) Lyapunov exponent $\lambda_1$, and (b) average frequency difference $\Omega_{\psi}$, for $\gamma=2.5$ and different values of the frequency modulation amplitude $k$. For $k=0$, the region of phase stability (as given by $\lambda_1 < 0$) and the region of permanent frequency entrainment (as given by $\Omega_{\psi}=0$) coincide; but for $k > 0$, the regions do not coincide: as $k$ is increased, the region with $\lambda_1 < 0$ is increased while the region with $\Omega_{\psi}=0$ is decreased. The graphs look very similar to those in the case of harmonic driving with bounded noise \cite{pikovsky2003}; therefore, investigation of time-variable dynamics for non-autonomously driven oscillators is necessary for an accurate understanding of the dynamical nature of the system. }
	\label{fig:noise-like}
\end{figure}

For our numerics in this section, we take $\theta_1(t)=\omega_1(t-\frac{k}{\omega_m}\cos(\omega_m t))$; so the frequency modulation $f(\omega_m t)$ is a sine wave $f(\omega_m t) = \sin(\omega_m t)$. Nonetheless, the results presented are just as valid for more general aperiodic slow modulation, and are demonstrated numerically to be true for aperiodic modulations in Sec.~\ref{sec:aperiodic}. We set $\omega_1 = 4$ and $\omega_m = 0.05$, except where stated otherwise, and we investigate the effect of the remaining free parameters $\gamma$, $\omega_0$, and $k$. We integrate the two-dimensional system~\eqref{eq:full_osc} with $r_p=1$ and $\epsilon=5$, except that for Fig.~\ref{fig:convergence} (showing synchronisation between solutions of \eqref{eq:system_polar_1} with different initial conditions) and Fig.~\ref{fig:noise-like} (showing long-term LE together with average frequency entrainment), we simply integrate \eqref{eq:system_polar_1}. All Lyapunov exponents, both long-term and time-localised, are computed following Benettin's canonical algorithm \cite{benettin1980a, benettin1980b}; for the time-localised LE, we use a moving average of the expansion coefficient. In~\eqref{eq:full_osc}, the radial LE at the limit cycle is equal to $-5$; therefore, since the maximum LE is greater than $-5$ in all our numerical experiments, it follows that this maximum LE corresponds to the phase dynamics defined by \eqref{eq:system_polar_1}. The same is also true of time-localised LE, at least after the first few moments needed for the trajectory to approach the limit cycle.

\begin{figure}[h]
	\centering
	\includegraphics[width=0.8\linewidth]{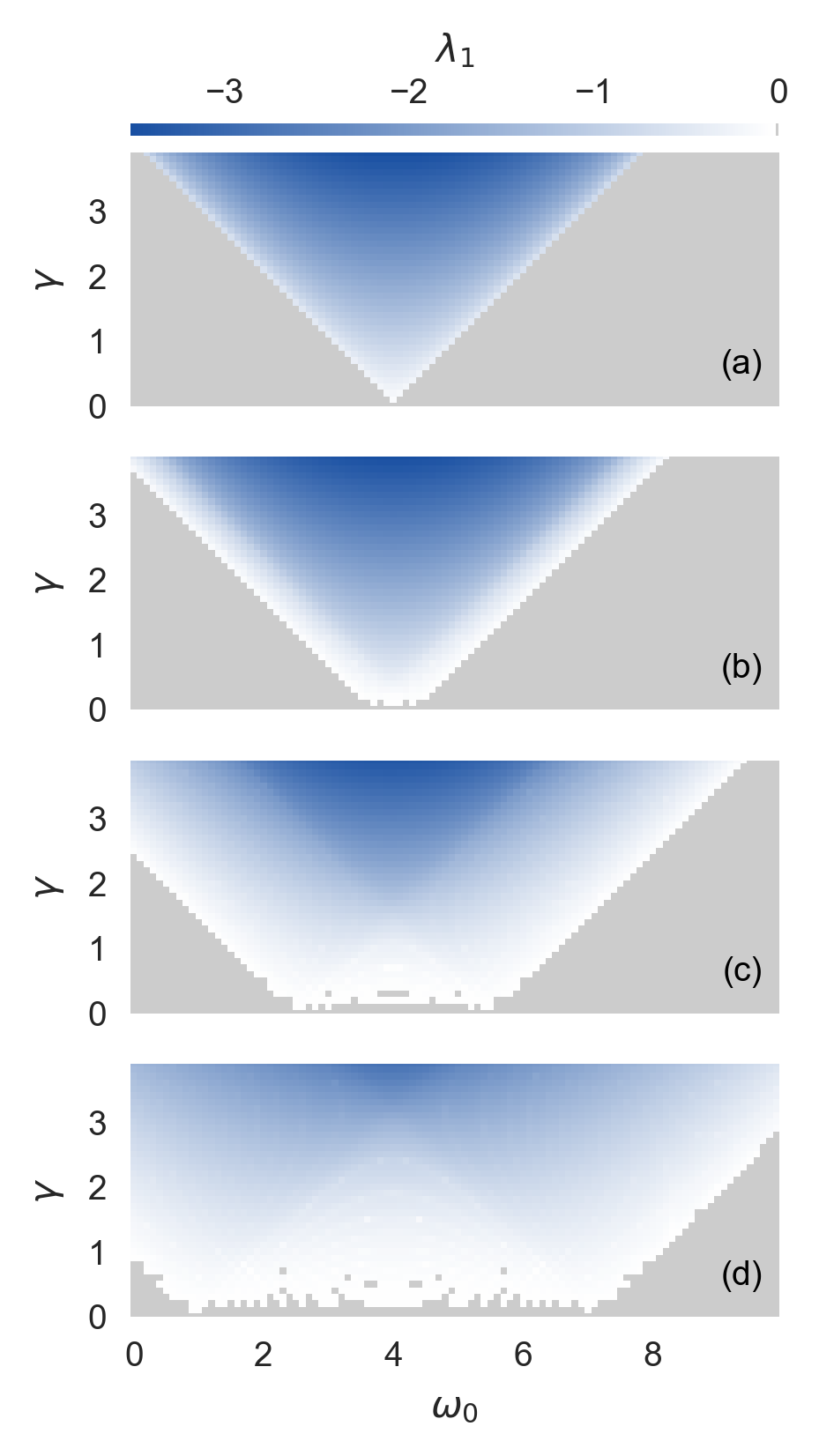}
	\caption{Numerically obtained long-term maximum Lyapunov exponent $\lambda_1$ over parameter space for \eqref{eq:full_osc}, with different $k$. The LE are computed over 5 cycles of the frequency modulation (about 630~s). In each case, 20 random initial conditions were taken from the square $[-1,1]\times [-1,1]$, and the average maximum LE over these trajectories is plotted. (a) $k=0$, (b) $k=0.1$, (c) $k=0.4$, (d) $k=0.8$. The Arnold tongue (shades of blue) is enlarged as $k$ increases. Grey represents zero values.}
	\label{fig:diagram_lyap}
\end{figure}

First, we investigate the long-term stability of the system, by means of the numerically computed maximum LE defined over a long time-window. Stability is indicated by a negative value for the LE. In Fig.~\ref{fig:diagram_lyap}, we see that there is an Arnold tongue (shades of blue) similar to that shown in Fig.~\ref{fig:k-fattening}(b). As illustrated in Fig.~\ref{fig:convergence}, solutions of Eq.~\eqref{eq:system_polar_1} synchronise with each other when the parameters lie in the Arnold tongue, but not when the parameters do not lie in the Arnold tongue. As shown in Fig.~\ref{fig:diagram_lyap}, the Arnold tongue is enlarged as the amplitude $k$ of the frequency modulation is increased.

%

\begin{figure*}[htb!]
	\centering
	\includegraphics[width=\linewidth]{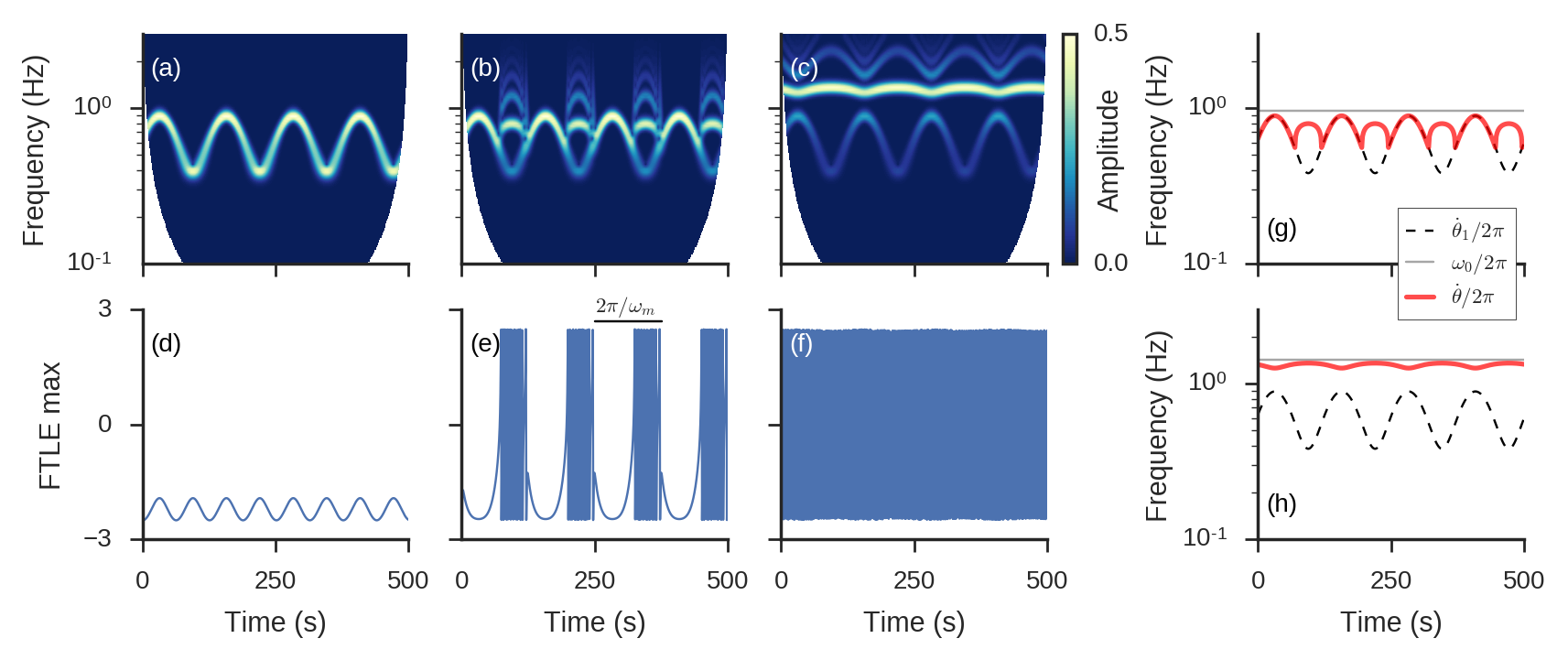}
	\caption{Analysis of the time-variable dynamical properties of a trajectory of \eqref{eq:full_osc}, in the three regions [see Fig.~\ref{fig:k-fattening}]; in (a-f), a random initial condition is taken from the square $[-1,1]\times [-1,1]$. (a-c)~Time-frequency representation (showing amplitude) of $\theta(t)$ obtained as the (unwrapped) polar angle of the trajectory of \eqref{eq:full_osc}, extracted using a continuous Morlet wavelet transform ($p=1$) with central frequency 3; and (d-f) maximum FTLE for the trajectory of \eqref{eq:full_osc} over a time-window of width $\tau = 0.1$~s for regions II, III, and I, from left to right. Parameters are set to $k=0.4$ and $\gamma = 2.5$; in (a,~d) $\omega_0=4$, in (b,~e,~g) $\omega_0=6$, and in (c,~f,~h) $\omega_0=9$. (a,~d) Region II exhibits frequency entrainment and a stable phase at all time.  (b,~e) Region III shows intermittent, but regular, epochs of frequency entrainment; the phase is stable on average over a long time. (c,~f) Region I, no frequency entrainment, and a FTLE rapidly oscillating around zero. (g,~h) Prediction (red) of the main observed frequency of the system over time, based on Eq.~\eqref{eq:main_freq} with values of $\Omega_\psi(t)$ taken from the $\Omega_{\psi}$ curve with $k=0$ in Fig.~\ref{fig:noise-like}(b), for (g) region III, and (h) region I. Interestingly, in region I, the main observed frequency oscillates in anti-phase with those of the driving frequency (dashed). Note that we here only predict the main frequency, and not the higher harmonics observed in (b,~c).}
	\label{fig:ftle_and_wt}
\end{figure*}

In other words, stability is induced by varying the frequency of the forcing over time. Quantitatively, we observe that the width of the Arnold tongue grows linearly with $k$.

While Fig.~\ref{fig:diagram_lyap} shows the long-time stability, region~III can only be distinguished and understood from the point of view of time-localised stability. The dynamics of Eq.~\eqref{eq:system_polar_1} is illustrated over time for the three regions in Figs.~\ref{fig:ftle_and_wt}(a-f), by time-frequency representation and by time-localised LE---namely, maximum LE defined over the time-window $[t,t+\tau]$ where $\tau$ is a fixed number. Here, we take $\tau=0.1$~s.

Region III is a region of intermittent synchronisation where trajectories alternate between epochs of time-localised stability and epochs of time-localised neutral stability; indeed, as the time $t$ evolves, the time-localised LE alternates between epochs where it is negative, and epochs where it oscillates with high frequency around an average value of zero, as is seen in Fig.~\ref{fig:ftle_and_wt}(e). Averaging over the total time yields a negative LE, meaning overall stability on average, even though the short-term stability is time-varying. Region~I is thus the only region with a long-term LE of zero, and this region decreases in size, which means that the region of stability increases.

The distinction between the three regions can be seen by looking at the time-frequency representation of a trajectory in each of these regions, as shown in Figs.~\ref{fig:ftle_and_wt}(a-c). In all three cases, the changing frequency of the driver is reflected in the frequency content of the driven oscillator. In Fig.~\ref{fig:ftle_and_wt}(a), representing region~II, the driving frequency is the only frequency present, as the frequency of the driven oscillator is entrained by that of the driving at all times. In Fig.~\ref{fig:ftle_and_wt}(c), representing region~I, we also see the natural frequency of the driven oscillator (cream, representing the highest amplitude), though slightly modulated by the driving. The fact that these two frequency modes are distinct shows that the driven oscillator's frequency is not entrained by the driving at any time. In Fig.~\ref{fig:ftle_and_wt}(b), representing region~III, we see the maximum-amplitude frequency mode overlapping the driving frequency at some times, but not at other times. The times of overlap are when the frequency of the driven oscillator is entrained by that of the driving, and the other times are when there is no frequency entrainment. Thus, in this region, we have intermittent frequency entrainment. Comparing (a,~b,~c) with (d,~e,~f) in Fig.~\ref{fig:ftle_and_wt}, we can see that in all three regions, absence of frequency entrainment coincides with time-localised LE that oscillate about $0$, while the occurrence of frequency entrainment coincides with time-localised LE that stay negative over a longer time-interval.

\begin{figure}[h]
	\centering
	\includegraphics[width=\linewidth]{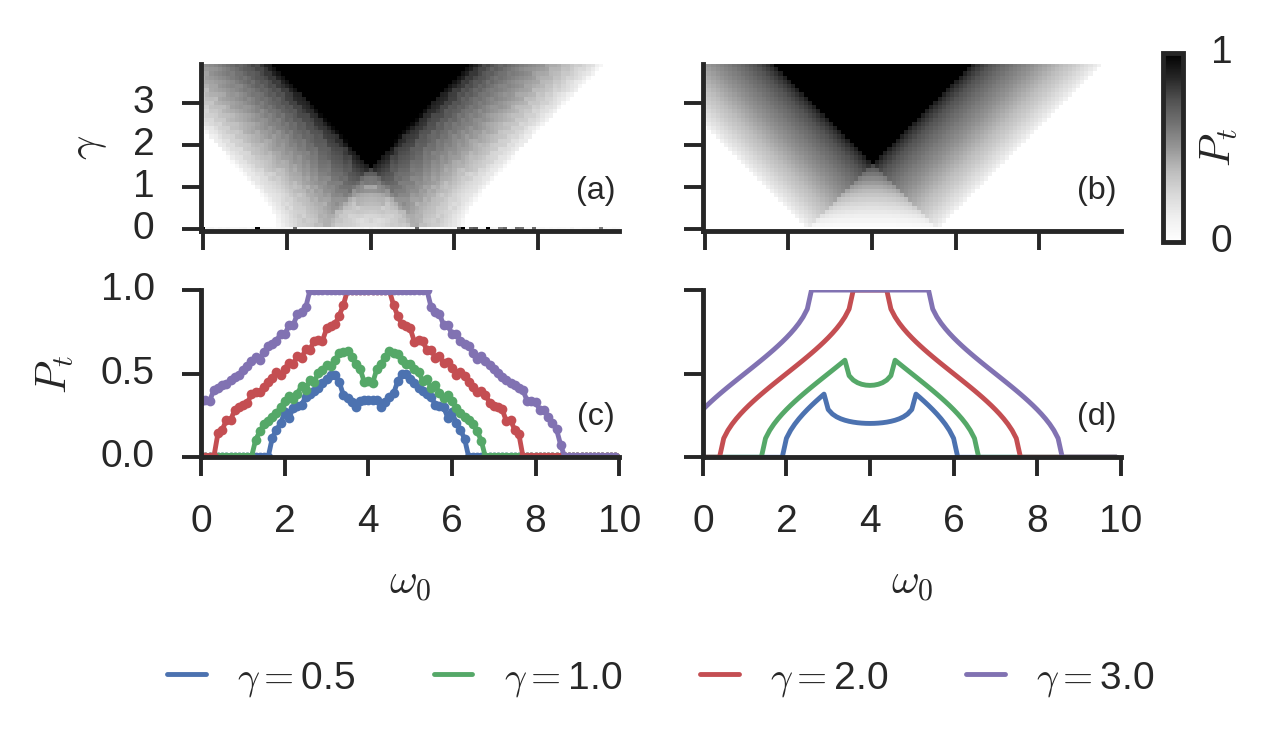}
	\caption{Analytical and numerical characterisation of the three regions based on the proportion of time spent in the stable regime, over $(\gamma, \omega_0)$-parameter space with $k = 0.4$. In (a,~c), $P_t$ is calculated numerically based on time-localised maximum LE (with window $\tau=0.1$~s) for 20 trajectories of \eqref{eq:full_osc} with random initial conditions in the square $[-1,1]\times [-1,1]$, over four cycles of the frequency modulation (about 500~s), and the result is averaged over the 20 trajectories. In (b,~d), the analytical result according to Eq.~\eqref{eq:analytical_pt} is shown. Plots~(c) and (d) show $P_t$ from (a) and (b) for selected $\gamma$ values. Three distinct regions appear clearly, respectively with $P_t$ values of zero (region~I), one (region~II), and in between zero and one (region~III). Analytical and numerical characterisations show good agreement.}
	\label{fig:stable_time}
\end{figure}

Now when investigating numerically the time-evolution of the time-localised LE, as in Fig.~\ref{fig:ftle_and_wt}(e) one can clearly distinguish between those time-intervals where the time-localised LE oscillates with high frequency around zero, and those time-intervals of length much greater than the periods of these aforementioned high-frequency oscillations during which the time-localised LE remains negative; and hence, one can numerically distinguish between the three regions. The proportion $P_t$ of time taken up by time-intervals where the time-localised LE remains negative is plotted in Figs.~\ref{fig:stable_time}(a,~c), across different parameter values. As in Figs.~\ref{fig:ftle_and_wt}(d--f), we expect $P_t=1$ in region~II, $0<P_t<1$ in region~III, and $P_t=0$ in region~I. We also plot in Figs.~\ref{fig:stable_time}(b,~d) the analytically obtained proportion $P_t$ of time for which the instantaneous vector field has a stable equilibrium. This is given by
\begin{widetext}
\begin{align}
P_t 
&=
\left\{\begin{array}{ll}
0, & (\text{A}) \text{ if } \gamma < |\Delta \omega (t)| \, \forall t, \\
1, & (\text{B}) \text{ if } \gamma > |\Delta \omega (t)| \, \forall t, \\
\frac{1}{\pi} \left[ \arcsin \left( \frac{\gamma - (\omega_0 - \omega_1)}{ \omega_1 k} \right)
- \arcsin \left( \frac{ -\gamma - (\omega_0 - \omega_1)}{ \omega_1 k} \right) \right], & (\text{C}) \text{ if } -\gamma \ge \Delta \omega_- \text{ and } \gamma \le \Delta \omega_+,\\
\frac{1}{\pi} \arcsin \left( \frac{\gamma - |\omega_0 - \omega_1|}{ \omega_1 k} \right) + \frac12, & (\text{D}) \text{ else},
\end{array} \right.
\label{eq:analytical_pt}
\end{align}
\end{widetext}
where $\Delta \omega_{\pm} = \omega_0 - \omega_1 (1 \mp k)$. The close resemblance between (a,~c) and (b,~d) helps confirm the validity of the numerical approach to distinguishing between the three regions. Fig.~\ref{fig:stable_time}(b) provides a quantitative picture for the qualitative skeleton shown in Fig.~\ref{fig:k-fattening}(b).

The average frequency difference $\Omega_{\psi} = \langle \dot{\psi} \rangle = \langle \dot{\theta} \rangle - \omega_1$ is a measure of the ``average frequency entrainment'' of the system \cite{pikovsky2003}. In the traditional autonomous case $k=0$ where the driving frequency is constant, nullity of $\Omega_\psi$ is equivalent to actual frequency entrainment. The quantity $\Omega_{\psi}$ is shown in Fig.~\ref{fig:noise-like}(b) for $\gamma = 2.5$ across different values of $k$. On Fig.~\ref{fig:noise-like}(a), the corresponding curves for the long-term Lyapunov exponent are displayed. Curves for $\Omega_{\psi}$ for $k>0$ are extremely similar to the case of driving with bounded noise $\xi(t)$, $\dot{\psi} = \Delta \omega + \gamma \sin(\psi) + \xi(t)$ (see \cite{pikovsky2003}). This will be explored in further detail in Sec.~\ref{sec:noise}. The similarity is due to the fact that only averages are considered, and time is forgotten. However, investigation of finite-time dynamics reveals that in region~III, the frequency difference alternates between epochs where it is zero and epochs where it is non-zero [as in Fig.~\ref{fig:ftle_and_wt}(g)]. To obtain this, we calculate the main observed frequency $\dot{\theta}_\textrm{main}/2\pi$ of trajectories using the slow modulation assumption, by taking
\begin{equation}
\dot{{\theta}}_\textrm{main}(t) = \dot{\theta}_1 (t) + \Omega_{\psi}(t)
\label{eq:main_freq}
\end{equation}
where $\Omega_{\psi}(t)$ is the $\Omega_\psi$-value associated with the autonomous differential equation $\frac{d}{ds}\psi(s) = \Delta \omega(t) + \gamma \sin \psi(s)$; results are plotted in Figs.~\ref{fig:ftle_and_wt}(g) and \ref{fig:ftle_and_wt}(h).

%

\subsection{Comparison between non-autonomous and noisy systems}
\label{sec:noise}

Experimental science essentially seeks to understand the underlying mechanics of a system that gives rise to the observed behaviour. Since the study of time-homogeneous dynamics (deterministic or noisy) is very well developed in comparison to the study of non-autonomous dynamics, there is a tendency to assume that for modelling purposes, the dynamics of a real-world system may be treated as statistically time-homogeneous. In this section we will illustrate, using our above-identified phenomenon of intermittent time-localised stability, how such a tendency may lead to the complete misidentification of some key aspect of the internal mechanics of a system.

There are various methods for analysing experimentally obtained time-series that are based on time-averaged properties of the time-series, such as power spectra. The theory of both deterministic autonomous dynamical systems and autonomous systems perturbed by stationary noise is well developed, and in particular it is well-known that adding noise to a system can create stability which was not present in the absence of noise, e.g.\ \cite{spagnolo1996, toral2001, newman2017, pikovskii1984, pikovsky2003}. Indeed, a one-dimensional phase oscillator model will almost invariably exhibit asymptotic stability of solutions when driven by stationary white noise \cite{malicet2017}. Therefore, when seeking to understand the mechanism by which a real-world system behaves robustly against unpredictable external perturbations, if one observes in a time-series of measurements from the system a power spectrum similar to that of some noisy model, and if moreover this noisy model is known to exhibit stability with negative Lyapunov exponents as a consequence of the noise, then naturally one may come to the conclusion that the real-world system under investigation is subject to a significant level of noise and that this noise plays the key role in causing stability.

However, our results for the deterministic system \eqref{eq:system_polar_1}--\eqref{eq:forcing_1_non-stationary} demonstrate that such a conclusion may be profoundly erroneous. The frequency modulation in \eqref{eq:forcing_1_non-stationary} may be an entirely deterministic process that is not subject to any significant levels of noise. This gives rise to the deterministic non-autonomous equation \eqref{eq:rotating_frame}, and we will illustrate that the time-averaged properties of \eqref{eq:rotating_frame} [with $f(\cdot)=\sin(\cdot)$] are very similar to those of a noisy counterpart
\begin{equation}
\dot \psi = (\omega_0 - \omega_1) + \gamma \sin \psi + \xi (t)
\label{eq:noise}
\end{equation}
where $\xi(t)$ is bounded noise. Physically, Eq.~\eqref{eq:noise} represents the phase difference under a model in which the driving frequency modulation is assumed to be noisy. The similarity that we shall illustrate between the time-averaged properties of \eqref{eq:rotating_frame} and \eqref{eq:noise} proves an important point: Since real-world systems are open and therefore subject to time-variability, one must examine temporally evolving dynamical properties of a system rather than just time-averaged properties, in order to account for the possibility that the mechanisms behind features of the observed behaviour are due to non-autonomicity. In the case of the system \eqref{eq:system_polar_1}--\eqref{eq:forcing_1_non-stationary}, in region~III, the mechanism behind stability is not stationary noise but deterministic intermittent frequency entrainment between driving and driven oscillators, arising from the slow variation of the driving frequency.


For simulations, dichotomous Markov noise $\xi(t)$, which switches between $\pm D$ at rate $\mu$ (with $\xi(0)=+D$), was used \cite{kim2006}. Nonetheless, it is expected that any bounded noise will show similar behaviour to that presented here \cite{pikovsky2003}; moreover, although the asymptotic properties of unbounded noise models exhibit a slightly different behaviour \cite{pikovsky2003}, any noise will effectively serve as bounded noise over typical physically relevant finite time-scales.

\begin{figure}[h]
\centering
\includegraphics[width=\linewidth]{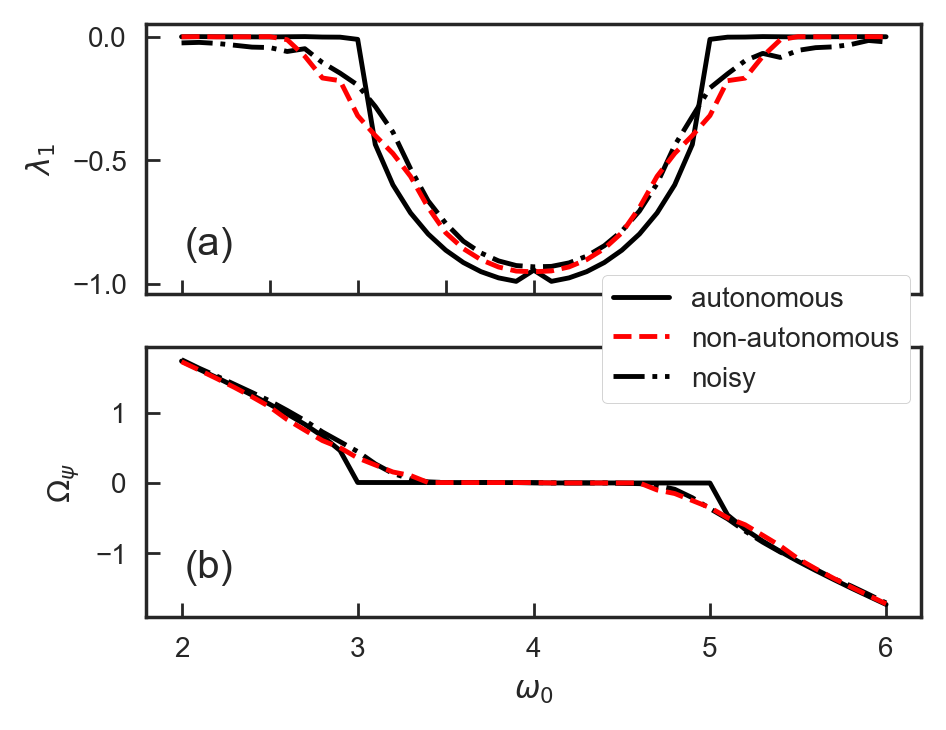}
\caption{Numerically obtained time-averaged stability properties for autonomous (plain black), non-autonomous (dashed red), and bounded noise (dotted-dashed black) driving, computed over about 1260~s (10 cycles of the periodic frequency modulation used for the non-autonomous case). For the autonomous and non-autonomous cases, Eqs.~\eqref{eq:auton_pd} and \eqref{eq:rotating_frame} respectively were integrated, with $\psi(0)=0$; for the noisy case, one sample path of $\xi(t)$ was generated, and Eq.~\eqref{eq:noise} was integrated with $\psi(0)=0$, using the same sample path $\xi(t)$ for all $\omega_0$ values. Parameters are set to $\omega_1=4$ and $\gamma=1$; for the non-autonomous case, $f=\sin(\cdot)$, $k=0.1$ and $\omega_m=0.05$, and for the noisy case $D=1.6$ and $\mu=10$. (a) Lyapunov exponent $\lambda_1$, and (b) average frequency difference $\Omega_{\psi}$. The non-autonomous and noisy cases, observed on average, present the same enlarging of the negative Lyapunov exponent region, and their $\Omega_{\psi}$ is almost exactly identical, including the plateau.}
\label{fig:noise_average}
\end{figure}

On average, the noisy and the non-autonomous systems will have properties as illustrated in Fig.~\ref{fig:noise_average} that look essentially the same. Indeed, both can be made to have an increased region for negative LE [see Fig.~\ref{fig:noise_average}(a)] and a smaller plateau for average frequency entrainment [see Fig.~\ref{fig:noise_average}(b)], as compared to the autonomous case given by $k=0$ or $D=0$.
%

\begin{figure*}[htb]
	\centering
	\includegraphics[width=\linewidth]{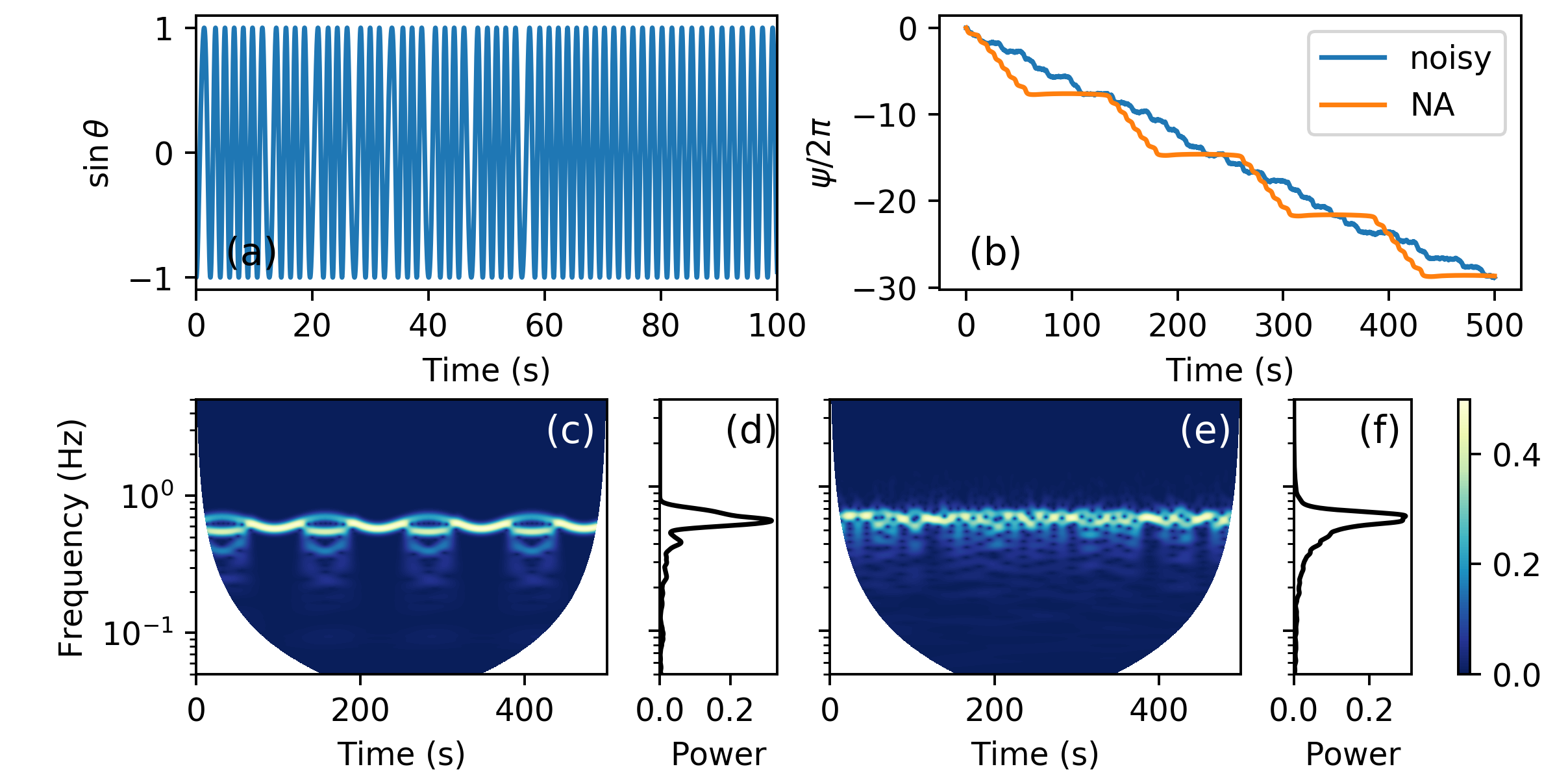}
	\caption{Dynamics of driven oscillator $\theta$ with (a,b,c,d) non-autonomous and (b,e,f) bounded noise driving; here $\theta$ evolves according to Eq.~\eqref{eq:system_polar_1} with $\theta_1(t)=\omega_1(t-\frac{k}{\omega_m}\cos(\omega_m t))$ in the non-autonomous case, and $\theta_1(t)=\omega_1t-\int_0^t \xi(s) \, ds$ in the noisy case, and $\theta(0)=\theta_1(0)$ in both cases. Parameters are set to $\omega_1=4$, $\gamma=1$, $\omega_0=3$; for the non-autonomous case $k=0.1$ and $\omega_m=0.05$, and for the noisy case $D=1$, $\mu=4$. First $\psi(t)$ is numerically obtained by integrating Eq.~\eqref{eq:rotating_frame} or \eqref{eq:noise} as appropriate, with $\psi(0)=0$, and then $\theta(t)$ is obtained by $\theta(t)=\psi(t)+\theta_1(t)$. (a) Sine of the driven phase $\theta$, in the non-autonomous setting. (b) Phase difference $\psi$ between the driving and driven oscillators, over time. The non-autonomous case presents epochs of phase-locking as seen by the regular plateaus, whereas the noisy phases' difference drifts without ever phase-locking to the driving, with an average velocity that is close to that of the non-autonomous curve. (c,~e) Time-frequency representation (showing power, i.e. square of the amplitude) for $\theta$ extracted using continuous Morlet wavelet transform (with $p=1$) with central frequency 3, and (d,~e) the associated time-averaged power. The main difference is the presence of intermittent frequency entrainment in the non-autonomous case. The average-power spectra are very similar, and do not clearly distinguish the two cases. Long-term LEs were also found to be negative in both cases: for the non-autonomous Eq.~\eqref{eq:rotating_frame} with $\psi(0)=0$, the LE over 500~s was about $-0.32$, and for the noisy Eq.~\eqref{eq:noise} with $\psi(0)=0$, the LE over 500~s was about $-0.24$.}
	\label{fig:noise_trajectories}
\end{figure*}

The noisy and the non-autonomous systems can, however, be distinguised based on their dynamics over time. This is illustrated in Fig.~\ref{fig:noise_trajectories} by trajectories and their time-frequency representation. In the non-autonomous case, one can see the regularly intermittent frequency entrainment between driving and driven phases in Fig.~\ref{fig:noise_trajectories}(c), where frequency entrainment corresponds to those times where the instantaneous power-frequency spectrum has only a single peak, and in Fig.~\ref{fig:noise_trajectories}(b), where frequency entrainment corresponds to the regular plateaus in the phase difference. By contrast, in the noisy case, the instantaneous power-frequency spectrum shown in Fig.~\ref{fig:noise_trajectories}(e) is significantly more bumpy around a peak that stays roughly fixed over time, and the phase difference in Fig.~\ref{fig:noise_trajectories}(b) looks like it is essentially drifting at all times. Despite these starkly visible differences in time-variable properties, the average power spectra as shown in Fig.~\ref{fig:noise_trajectories}(d,~f) are reasonably similar to each other.



\subsection{Aperiodic modulation}
\label{sec:aperiodic}

\begin{figure}[h]
	\centering
	\includegraphics[width=\linewidth]{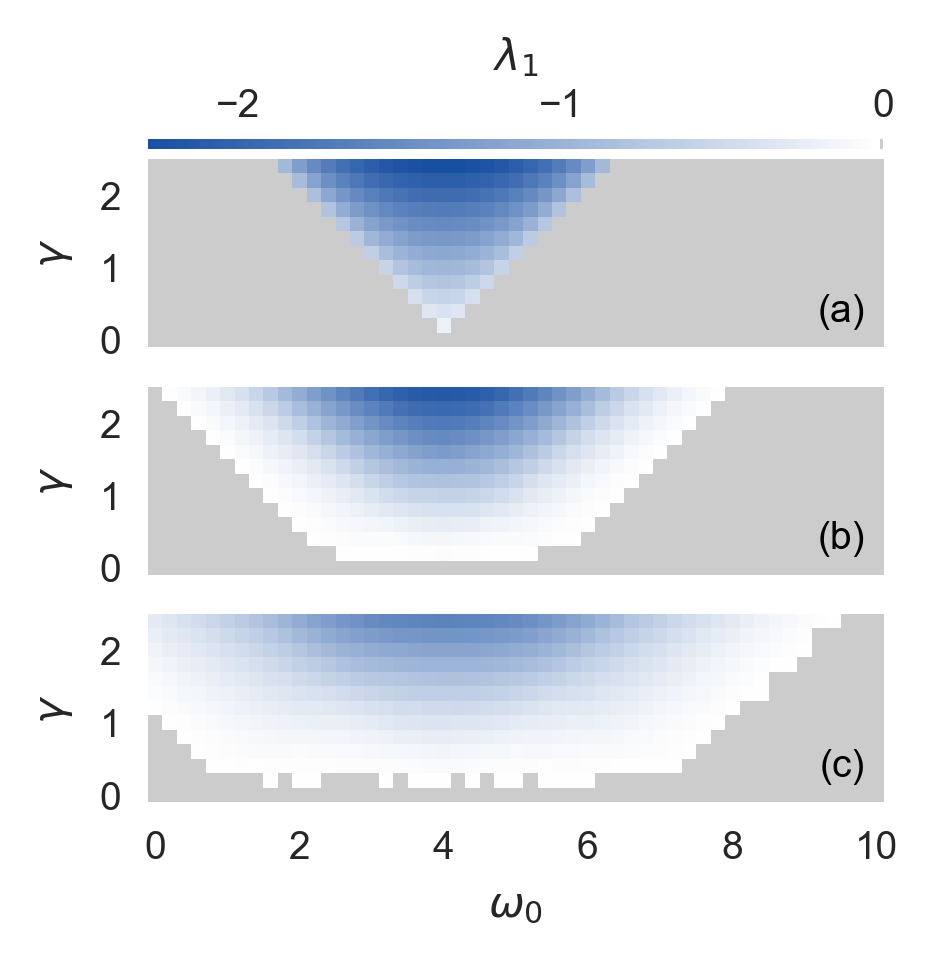}
	\caption{Numerically obtained long-term maximum Lyapunov exponent $\lambda_1$ over parameter space for \eqref{eq:full_osc} with $\theta_1(t)=\omega_1(t + \frac{0.5k}{\omega_m}(\sin(\omega_m t) + \frac{4}{\pi}\sin(\omega_m \pi t / 4)))$. Here, as in Sec.~\ref{sec:numerics}, $\omega_1=4$, $\omega_m=0.05$, $r_p=1$ and $\epsilon=5$. The LE are computed over 5 cycles of the frequency modulation (about 630~s). In each case, 20 random initial conditions were taken from the square $[-1,1]\times [-1,1]$, and the average maximum LE over these trajectories is plotted. (a)~$k=0$, (b)~$k=0.4$, (c)~$k=0.8$. The Arnold tongue (shades of blue) is enlarged as $k$ increases. Grey represents zero values.}
	\label{fig:quasi-periodic}
\end{figure}

We now consider numerically a quasi-periodic frequency modulation function $f(\cdot)$ given by
\begin{equation}
f (t) = 0.5[\cos(\omega_m t) + \cos(\omega_m \pi t / 4)].
\label{eq:quasi-periodic}
\end{equation}
As expected, and shown in Fig.~\ref{fig:quasi-periodic}, the enlargement of the Arnold tongue holds. Moreover, more quantitatively, the results shown in Fig.~\ref{fig:quasi-periodic}(b,~c) are almost identical to those shown in Fig.~\ref{fig:diagram_lyap}(c,~d) respectively. This is because $f(\cdot)$ oscillates throughout the interval $[-1,1]$ in both cases, and therefore Eqs.~\eqref{eq:qualitative_regions1}--\eqref{eq:qualitative_regions3} for the three different regions still hold.

\section{Higher-dimensional cases}
\label{sec:higher-dimensional}

In the above section, we showed analytically, and confirmed numerically, that enlargement of the stability region will always occur in the simple one-dimensional case. Nonetheless, the phenomenon of stabilisation by slow variation of the driving frequency, and the phenomenon of intermittent synchronisation under such variation of the driving frequency, may be found in a broader class of systems. To illustrate the more general scope of the stabilisation phenomenon, we illustrate it numerically in non-linear driven oscillators. We consider three cases: first, a typically forced van der Pol (vdP) oscillator; second, a vdP oscillator with the phase driven via diffusive coupling; and finally, a typically forced Duffing oscillator. All three cases are investigated with non-autonomous driving $\theta_1(t)=\omega_1(t-\frac{k}{\omega_m}\cos(\omega_m t))$, with $\omega_m=0.02$. Long-term LE are computed over 10~cycles of the frequency modulation (about 3140~s).

\subsection{Typically forced van der Pol}

We consider a vdP oscillator that is \emph{directly} forced by the external phase oscillator $\theta_1(t)$, so that the vdP oscillator satisfies the differential equation
\begin{equation}
\ddot x = \epsilon (1 - x^2)  \dot x - \omega_0 ^2 x + \gamma \sin(\theta_1(t)).
\label{eq:forced_vdp}
\end{equation}

\begin{figure}[h]
	\centering
	\includegraphics[width=\linewidth]{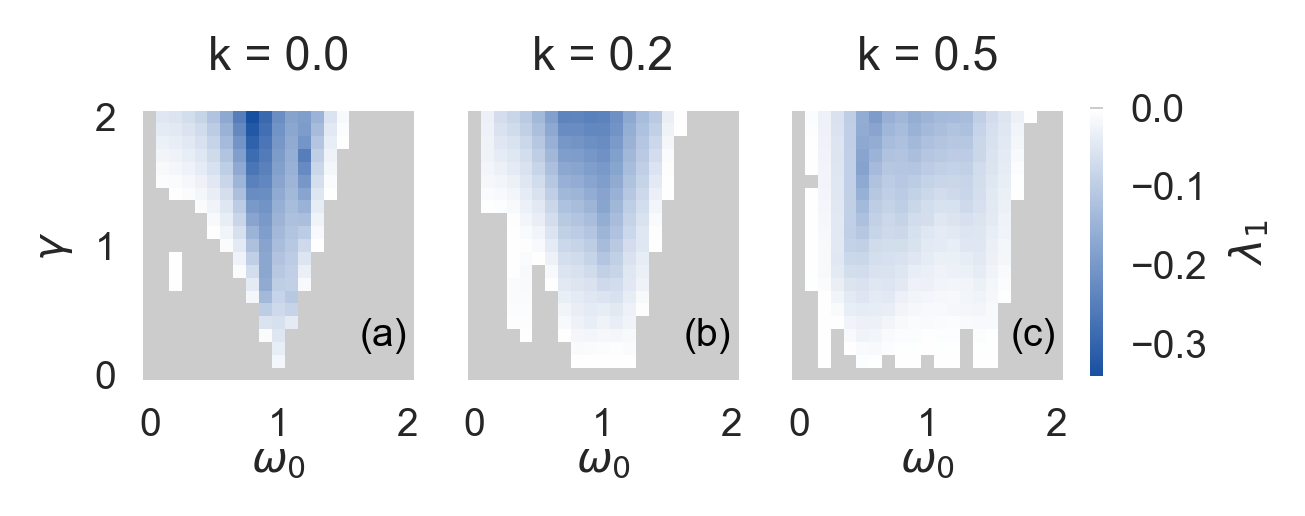}
	\caption{Numerically obtained long-term maximum Lyapunov exponent $\lambda_1$ over parameter space for the forced weakly non-linear vdP oscillator [see Eq.~\eqref{eq:forced_vdp}], with $\epsilon = 0.1$, $\omega_1 = 1$, and $\omega_m = 0.02$, for different amplitude of frequency modulation $k$. In each case, 20 random initial conditions $(x(0),\dot{x}(0))$ were taken from the square $[-1,1]\times [-1,1]$, and the average maximum LE over these trajectories is plotted. The negative LE region (blue shades) increases as $k$ is increased. Grey represents zero values.}
	\label{fig:forced_vdp_tongue}
\end{figure}

For investigation of LE, we treat this as a first-order equation in $(x,y)$-space with $\dot{x}=y$ (and numerically integrate it as such). The long-term maximum LE is shown over parameter space in Fig.~\ref{fig:forced_vdp_tongue}. The region with negative LE increases with the amplitude $k$ of the frequency modulation, showing that the enlargement of the negative LE region still holds
in this non-linear case.

\begin{figure}[h!]
\centering
\includegraphics[width=\linewidth]{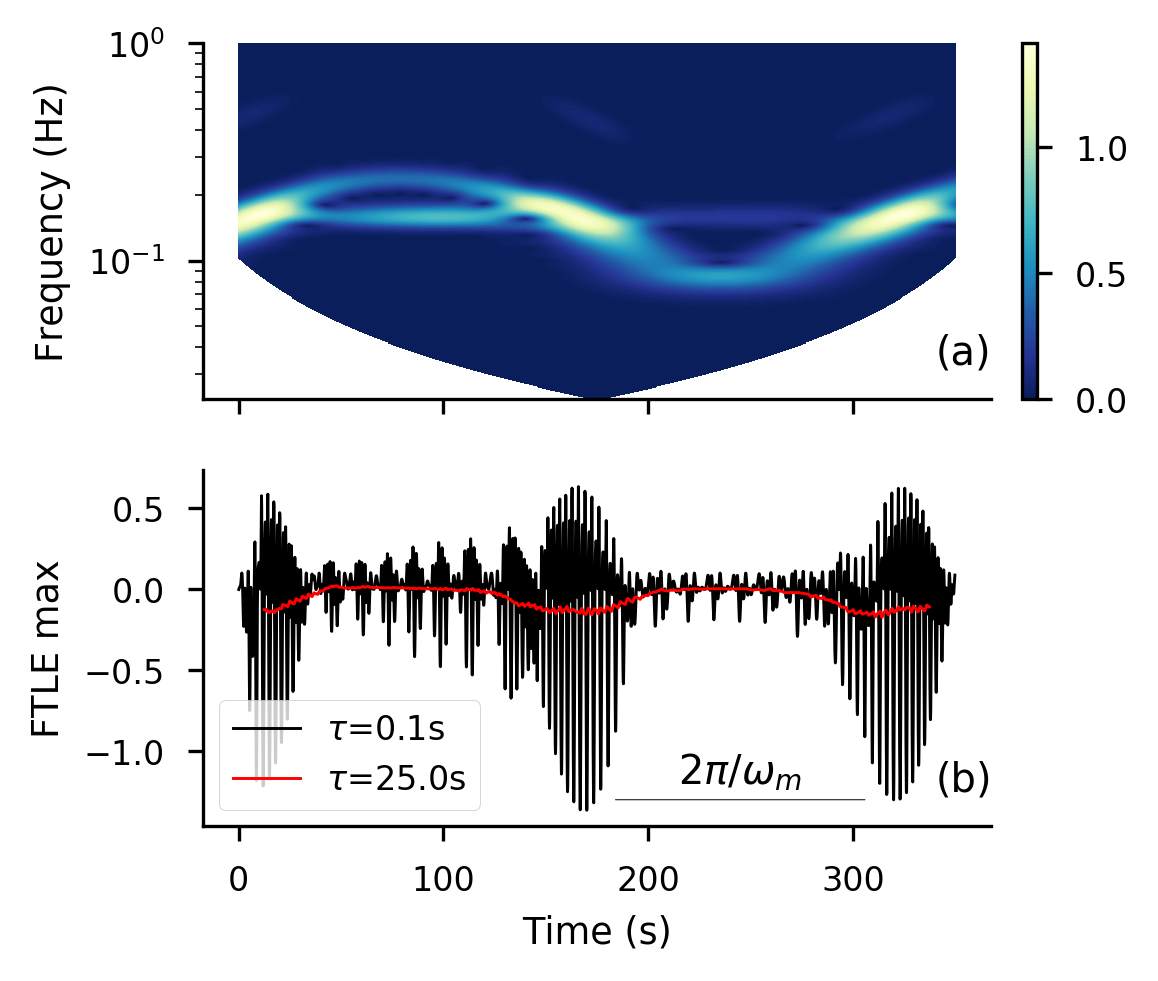}
\caption{Intermittency in the typically forced vdP system \eqref{eq:forced_vdp}. Parameters are set to $k=0.5$, $\gamma=1$, $\omega_0=1$, $\epsilon=0.1$, $\omega_m=0.02$, $\omega_1=1$; results are for a random initial condition $(x(0),\dot{x}(0))$ taken from the square $[-1,1]\times [-1,1]$. (a) Time-frequency representation (showing amplitude) of $x(t)$, extracted using a continuous Morlet wavelet transform ($p=1$) with central frequency 2. (b) Shorter-time-window FTLE max, with window length 0.1~s, is shown in black, and longer-time-window FTLE max, with window length 25~s, is shown in red. The longer-time-window FTLE alternates between epochs where it is negative, and epochs where it is zero. These epochs of negative values coincide with those epochs where, in (a), there appears to be a single main peak in the instantaneous power-frequency spectrum.}
\label{fig:phase_forced_vdp_series}
\end{figure}

Moreover, consideration of time-localised LE, as shown in Fig.~\ref{fig:phase_forced_vdp_series}(b), suggests that for some parameter values we have intermittency between epochs of stable dynamics and epochs of neutrally stable dynamics. Unlike in Fig.~\ref{fig:ftle_and_wt}(e), the stable epochs are not characterised by negativity of LE defined over a very short sliding window, but rather over a suitably longer sliding window. This is because, if one were to freeze the driving frequency $\dot{\theta}_1$ at any moment in time during such an epoch of stable dynamics, the solution of the resulting periodic differential equation \eqref{eq:forced_vdp} would not converge to a fixed point but most likely to a stable periodic orbit, for parts of which the vector field is locally contractive and other parts not, with contraction on average over each period. Hence, the intermittency is demonstrated most clearly by taking time-localised LE over a wider moving time-window, whose width is likely to incorporate several periods of the aforementioned stable periodic orbit. In Fig.~\ref{fig:phase_forced_vdp_series}(b), we see quite clearly (in red) the alternation between plateaus of zero time-localised LE and epochs where the time-localised LE dips to become negative. In the time-frequency representation shown in Fig.~\ref{fig:phase_forced_vdp_series}(a), during the epochs of zero time-localised LE, the power is shared mostly between two distinct peaks in the instantaneous spectrum, but during the epochs of negative time-localised LE, virtually all the power is concentrated around a single peak. As in Fig.~\ref{fig:ftle_and_wt}(e), this suggests that instantaneous $1:1$~frequency entrainment is taking place during the epochs of negative time-localised LE, but not the epochs of zero time-localised LE, and so overall the system exhibits intermittent synchronisation.


\subsection{Diffusively forced van der Pol}

\begin{figure}[h!]
	\centering
	\includegraphics[width=\linewidth]{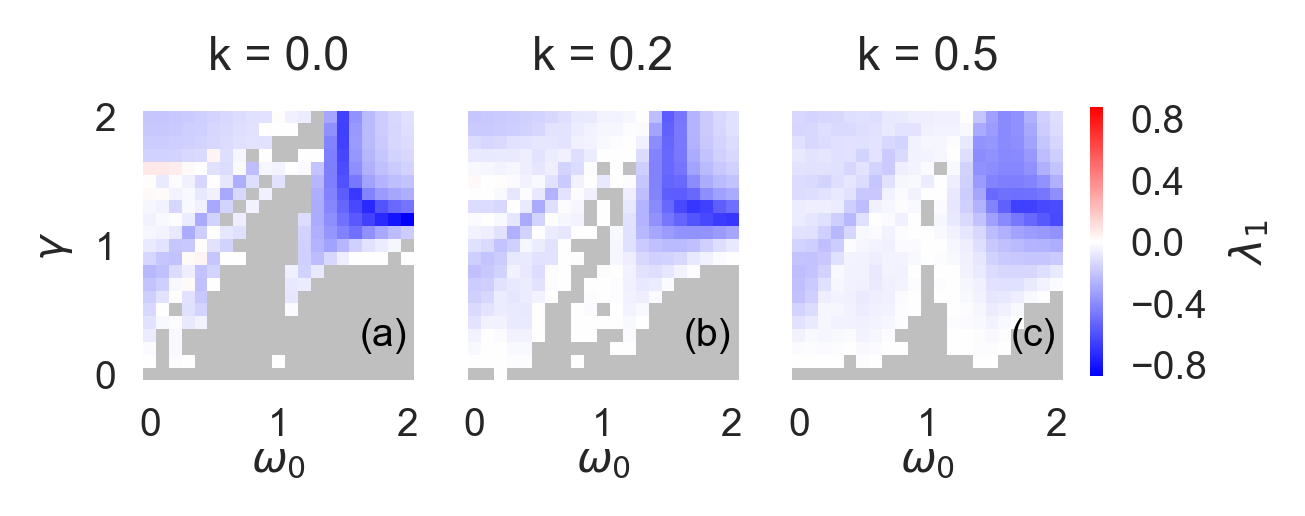}
	
	\caption{Numerically obtained long-term maximum Lyapunov exponent $\lambda_1$ over parameter space for the diffusively phase forced van der Pol oscillator [see Eq.~\eqref{eq:diff_phase_forced_vdp}], with $\epsilon = 0.1$, $\omega_1 = 1$, and $\omega_m = 0.02$, for different amplitude of frequency modulation $k$. In each case, 20 random initial conditions $(x(0),\dot{x}(0))$ were taken from the square $[-1,1]\times [-1,1]$, and the average maximum LE over these trajectories is plotted. The stability region (shades of blue) is enlarged as $k$ increases, and chaotic (red) points in (a) are turned stable (shades of blue) in (c). Grey represents zero values.}
	\label{fig:phase_diff_forced_vdp_tongues_k}
\end{figure}

We take the polar coordinate representation of the unforced vdP oscillator [Eq.~\eqref{eq:forced_vdp} without $\gamma\sin(\theta_1(t))$] as a first order equation in $(x,\dot{x})$-space, and we now drive the angular component with a diffusive coupling:
\begin{equation}
\begin{aligned}
\dot r =& (1 - \omega_0^2) r \cos \theta \sin \theta + \epsilon (1 - r^2 \cos^2 \theta) r \sin ^2 \theta, \\
\dot \theta =& \epsilon (1 - r^2 \cos^2 \theta) \sin \theta \cos \theta - \omega_0^2  \cos^2 \theta - \sin^2 \theta \\
& + \gamma \sin (\theta - \theta_1(t)).
\end{aligned}
\label{eq:diff_phase_forced_vdp}
\end{equation}
The long-term maximum LE is shown over parameter space in Fig.~\ref{fig:phase_diff_forced_vdp_tongues_k}. Increasing $k$ reduces both the region of neutral stability and the very small region of chaos, while the region of stability grows.
	
\subsection{Forced and coupled Duffing oscillator}

\begin{figure}[h]
	\centering
	\includegraphics[width=\linewidth]{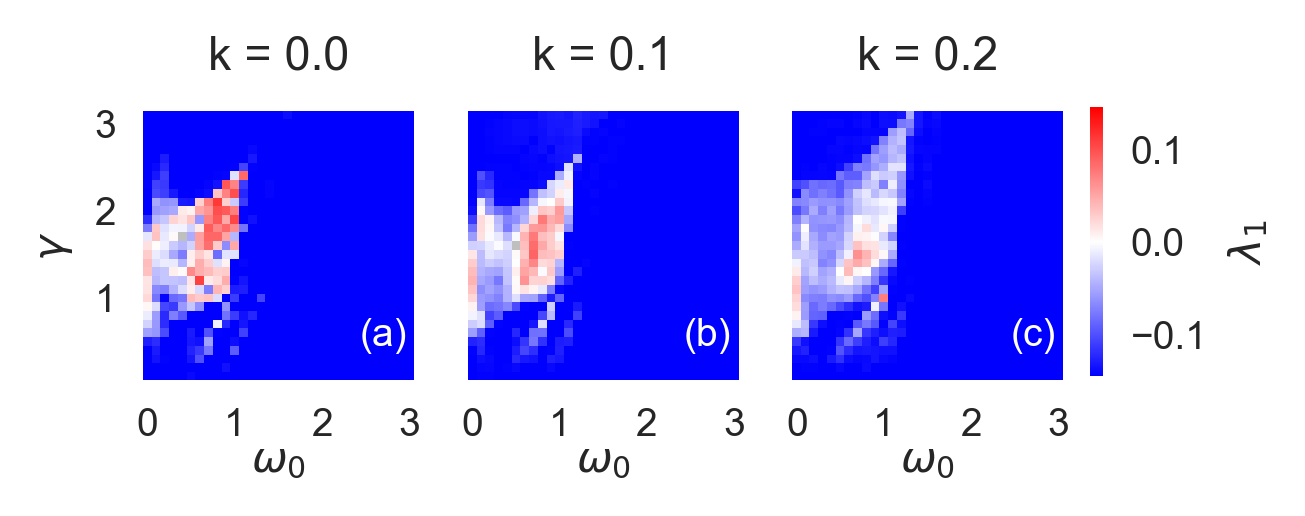}
	\caption{Numerically obtained long-term maximum Lyapunov exponent $\lambda_1$ over parameter space for the coupled and forced Duffing oscillator \eqref{eq:forced_duffing}, for different amplitude of frequency modulation $k$. In each case, 20 random initial conditions $(x(0),\dot{x}(0),x_1(0),\dot{x}_1(0))$ were taken from $[-1,1]^4$, and the average maximum LE over these trajectories is plotted. The region of chaotic behaviour (red) is reduced as many points are turned stable (shades of blue) as $k$ is increased. Grey represents zero values.}
	\label{fig:forced-duffing-k-varied}
\end{figure}

Here, we consider two Duffing oscillators, $x$ and $x_1$, unidirectionally diffuively coupled with strength $g_d$ so that $x_1$ drives $x$. Additionally, the driven Duffing oscillator is directly forced with external non-autonomous driving:
\begin{equation}
\begin{aligned}
\ddot x &= - \delta \dot x - \omega_0^2 x - \beta x^3 + g_d (x - x_1) + \gamma \cos(\theta_1(t)), \\
\ddot x_1 &= - \delta \dot x_1 - \omega_0^2 x_1 - \beta x_1^3 ,
\label{eq:forced_duffing}
\end{aligned}
\end{equation}
We fix parameters $\delta = 0.3$, $\beta = 0.1$, $\omega_1=1.2$, $\omega_m=0.02$, and coupling strength $g_d = 0.5$. For investigation of LE, we treat the system as a first-order equation in $(x,y,x_1,y_1)$-space with $\dot{x}=y, \, \dot{x}_1=y_1$ (and numerically integrate it as such). The long-term maximum LE is shown over parameter space in Fig.~\ref{fig:forced-duffing-k-varied}. As $k$ is increased, the chaotic region essentially decreases, giving way to either stability or neutral stability. The stability region does not strictly increase, as parts of the stability region become neutrally stable as $k$ is increased; nonetheless, the phenomenon is still observed that for various fixed values of all the parameters other than $k$, increasing $k$ has the effect of turning chaos into stability. From a control point of view, if the region of interest in parameter space is the chaotic region, one can stabilise the dynamics by adding time-variation to the forcing frequency.

\section{Discussion}
\label{sec:discussion}

The work was motivated by real systems that exhibit dynamics with time-varying frequencies and are stable against external perturbation
\cite{Stefanovska2000, Mcguinness2004, suprunenko2013, Tass:98,Varela2001, desaedeleer2013, Stefanovska2000, hramov2006,lancaster2016}. Surprisingly, not much analytical work has been carried out on such systems, and most of the work that has been carried out has used noisy driving \cite{Stratonovich1967, spagnolo1996, newman2017} as the foundation of the model, or noise consisting of impulses at random times \cite{pikovsky2003,pikovskii1984}. In these studies, it was shown that noise can create and increase stability. Outside of a stochastic approach, the only other way to incorporate time-variability is to model the system as a deterministic non-autonomous dynamical system. However, not much analytic theory of non-autonomous systems has been developed yet. The problem is additionally complicated by the fact that an asymptotic approach does not give the full picture as the evolving dynamics over shorter time-scales is missed. As illustrated in this work, changes in dynamical behaviour over shorter time-scales are of crucial importance in the types of systems considered. For if they had not been considered in this work, the phenomenon of intermittent synchronisation would have been missed. The work in this paper has provided a key new insight into systems subject to time-varying influences, by identifying the phenomenon of intermittent synchronisation and the region in parameter space where it occurs, and thereby showing the enlargement of the Arnold tongue. This insight also has potential for being the foundation of future methods to induce stability in complex or other systems; the fact that non-autonomous driving allows for average stability to be achieved without the need to maintain frequency entrainment at all times may be of significant advantage.

The basic adiabatic reasoning underlying our analytical approach is the same as that employed and investigated in \cite{jensen2002}. It is this reasoning that has led us to our new discovery that increasing time-variability inherently induces stability in phase oscillators. We have also employed numerical tools to visualise time-localised dynamics as derived by this adiabatic reasoning, namely time-localised LEs as in Figs.~\ref{fig:ftle_and_wt}(d-f) and time-frequency representation as in Figs.~\ref{fig:ftle_and_wt}(a-c). We hypothesise that for a time-frequency representation applied to experimental data, a result resembling Fig.~\ref{fig:ftle_and_wt}(b) could be a signature of intermittent synchronisation. We also investigated the slow variation assumption in higher-dimensional systems, and numerically illustrated the creation of stability as the amplitude of variation is increased, as well as the occurrence of intermittent synchronisation. In this way, we showed that the phenomena of stabilisation and intermittent synchronisation under slow variation of the driving frequency occur more broadly than just in the case of phase oscillators.

Chronotaxic systems have been introduced to model the distinctive feature of real-life oscillatory systems, that they are able to keep their time-varying dynamics resistant to external perturbations \cite{suprunenko2013, suprunenko2014a, suprunenko2014b}. Chronotaxic systems were defined in previous works by the necessary condition that a time-dependent attractor exists, and that trajectories in its close vicinity always move closer to it \cite{suprunenko2013}, or alternatively just by the existence of a positively invariant time-varying region in which the dynamics is always contracting \cite{suprunenko2014b}. However, it seems reasonable to expect that in real life, there exist stable oscillatory systems for which no trajectory is always instantaneously locally attractive. Instead, in contrast to currently existing definitions of chronotaxicity, an intermittent synchronisation phenomenon such as identified in this work would give rise to stability on average. Thus, we have broadened the definition of chronotaxic systems, increasing its potential for effectively modelling and understanding real-life systems, which are non-isolated and therefore continuously subjected to time-varying external influences.

\bigskip

\section{Summary}
\label{sec:summary}
We have shown that driving a phase oscillator with an arbitrary slowly varying frequency always induces stability: the larger the amplitude of the frequency modulation, the larger the stabilty region. We have furthermore shown numerically that this phenomenon occurs in more complex cases where the driven oscillator is higher-dimensional and non-linear, hinting at the wider scope and importance of the effect at hand. We have even shown numerically that chaotic regions in parameter space can be made stable by the same mechanism. If only the quantities $\lambda_1$ and  $\Omega_{\psi}$, which describe time-averaged properties of the system, are considered, the system looks the same as in the case of driving with noise. However, in reality, our fully deterministic example exhibits some time-localised frequency entrainment, whereas none is exhibited in the case of driving with bounded noise. It is therefore clear that the non-autonomous deterministic system could be misinterpreted as a noisy one if only time-averaged quantities are considered.
%

The enlarged stability region makes time-variable driving very suitable for real-world modelling and for engineering, where a controlled adjustment of the frequencies is often of key importance. We believe that this type of model will find applications in many fields, including physics, biology, medicine, and climate dynamics.


\section*{Acknowledgment}

We thank Arkady Pikovsky and Martin Rasmussen for valuable discussions, as well as Phil Clemson, Bastian Pietras and Tomislav Stankovski, for useful comments on the manuscript. We especially thank Yevhen Suprunenko for his invaluable help and for the useful discussions that we had. This work has been funded by the EU's Horizon 2020 research and innovation programme under the Marie Sk\l{}odowska-Curie grant agreement No~642563, and the EPSRC grant EP/M006298/1 \emph{A device to detect and measure the progression of dementia by quantifying the interactions between neuronal and cardiovascular oscillations}.

\bibliographystyle{apsrev4-1}


%

\end{document}